\documentclass[aip,apl,twocolumn,10pt,showpacs,longbibliography]{revtex4-1}
\pdfoutput=1
\usepackage{graphicx}
\usepackage{amsfonts}
\usepackage{amsmath}
\usepackage{natbib}

\begin{document}

\title{Error detection in a tunable-barrier electron pump}

\author{S.~P.~Giblin,$^{1}$ P.~See,$^{1}$ J.~D.~Fletcher,$^{1}$ T.~J.~B.~M.~Janssen,$^{1}$
J.~P.~Griffiths,$^{2}$ G.~A.~C.~Jones,$^{2}$ I.~Farrer,$^{2}$ and M.~Kataoka$^{2}$}

\affiliation{$^{1}$ National Physical Laboratory, Hampton Road, Teddington, Middlesex TW11 0LW, United Kingdom }

\affiliation{$^{2}$ Cavendish Laboratory, University of Cambridge, J J Thomson Avenue, Cambridge CB3 0HE, United Kingdom }
\email[stephen.giblin@npl.co.uk]{Your e-mail address}

\date{\today}

\begin{abstract}
We measure the average number of electrons loaded into an electrostatically-defined quantum dot (QD) operated as a tunable-barrier electron pump, using a point-contact (PC) charge sensor 1 micron away from the QD. The measurement of the electron number probes the QD loading dynamics even in the limit of slow gate voltage rise-times, when the pumped current is too small to measure. Using simulations we show that, with optimised QD-PC coupling, the experiment can make single-shot measurements of the number of electrons in the QD with sufficiently high fidelity to test the error rate of the electron pump with metrological precision.
\end{abstract}

\pacs{1234}

\maketitle

\section{Introduction}

The semiconductor tunable-barrier electron pump \cite{blumenthal2007gigahertz,kaestner2008single,fujiwara2008nanoampere,jehl2013hybrid,rossi2014accurate} is a promising candidate for a primary realisation of the ampere in a re-defined SI system based on fundamental constants. The pump has so far demonstrated a current accuracy of 1.2 parts per million for a relatively high current of 150 pA \cite{giblin2012towards}. This test was performed by comparing the pump current with primary standards derived from the Josephson effect and quantum Hall effect, and more accurate current comparison tests are feasible, for example by using a Cryogenic Current Comparator to amplify the relatively small pump current \cite{feltin2003progress}. However, mesoscopic charge sensing techniques make it possible to perform an entirely different type of accuracy test, in which electron pumping events (or alternatively, errors in pumping events) are detected one at a time. Tests of this type provide a more robust demonstration of the electron transfer accuracy than average current measurements, and are required if the pump is to be used in a metrological-triangle type experiment \cite{keller2008current}. They are also not limited by the accuracy with which the pump current can be compared with primary standards. To demonstrate the desired metrological error rate of $1$ in $10^8$ or less, with reasonable statistical confidence, it should be possible to perform at minimum $10^8$ pump-detect operations within a reasonable experimental time-scale (for fundamental metrology experiments) of $1-2$ days.

One approach to measuring pump errors, is to pump the electrons on and off a mesoscopic island which is capacitively coupled to a charge sensor such as a single-electron transistor (SET) or quantum point contact QPC). This technique, using SET detectors, has been successfully employed to measure the error rate of multi-junction metallic pumps \cite{martinis1994metrological,keller1996accuracy,camarota2012electron}, pumping at relatively low frequencies $\lesssim 10$~MHz. In one experiment an error rate of $15$ errors in $10^9$ pump operations was demonstrated \cite{keller1996accuracy}. This result was based on recording $\sim 30$ errors in data sets $\sim 500$ seconds long incorporating $\sim 2.5\times 10^9$ pump operations. In this type of 'shuttle' experiment, the charge sensing bandwidth only needs to accommodate the error rate and not the overall pump rate. Recently, two experiments have applied the island-pumping concept to measuring the error rate of the tunable-barrier semiconductor pump \cite{fricke2013counting} and turnstile \cite{yamahata2014accuracy}. These experiments measured error rates of $0.009$\cite{fricke2013counting} and $0.0001$\cite{yamahata2014accuracy}, transferring $1$ and $2$ electrons per cycle respectively at a rate slow enough (in the range $10-100$~Hz) to detect every electron and therefore accumulate full counting statistics of the transferred charge. The experiment of Fricke et al, rather than pumping electrons in two directions on and off one island, employed multiple series pumps to move electrons in one direction, and the resulting small $\sim 5$~aA DC current, with accompanying counting statistics, can be treated as a self-contained representation of the ampere \cite{fricke2014self}. This concept can, in principle, be scaled-up to count errors at GHz pump frequencies at the expense of quite formidable complexity, using 5 pumps in series, and 4 islands in between the pumps each monitored by a high-bandwidth RF-SET \cite{fricke2014self}.

In this work, we demonstrate a new approach to measuring the pumping errors, in which the number of electrons loaded into the pump is measured directly, during a pause in the pump cycle, utilizing a nearby QPC charge sensor. The charge sensing island is now the pump quantum dot (QD) itself. QPC sensors have been used to probe QD electron occupation in many types of experiment, for example to measure equilibrium tunneling dynamics between a QD and leads \cite{vandersypen2004real}, spin dynamics via spin-charge conversion \cite{elzerman2004single} and non-equilibrium tunneling out of a many-electron dot \cite{cooper2000direct}. In our experiment, we use the QPC to measure the average number of electrons loaded into the QD and held in a non-equilibrium state above the Fermi level. We show very good agreement between current measurements which probe the average number of electrons pumped through the QD, and QPC measurements which probe the average number of electrons trapped in the QD. Furthermore, the QPC signal can probe the QD loading dynamics when the pumping is too slow to generate a measurable current. Finally, we use simulated data to show that our method can perform a single-electron metrological accuracy test of the tunable-barrier electron pump with quite modest and achievable improvements in the QD-PC coupling sensitivity.

\begin{figure}
\includegraphics[width=8.5cm]{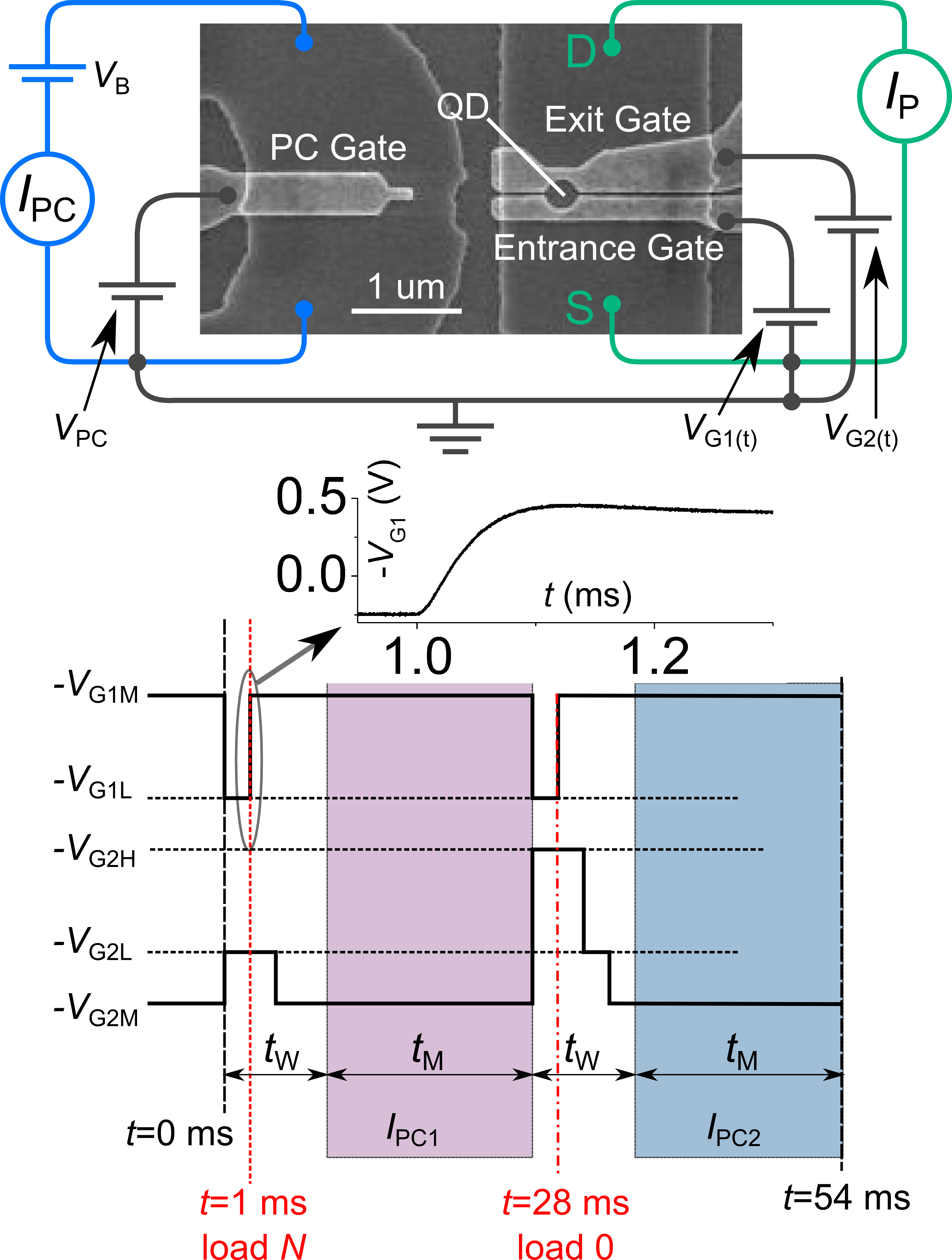}
\caption{\textsf{(a): SEM image of device and electrical connections to measurement circuit. (b):schematic showing time dependence of entrance gate (upper line) and exit gate (lower line, offset on y-axis for clarity) voltages supplied to the pump gates when the device is operated in electron trapping mode. The negative of the gate voltage is shown so that the traces indicate the barrier height under the gates. The shortest step time is $1$~ms. The inset shows a representative measurement of the gate voltage rise time, using a real-time oscilloscope.}}
\label{fig:fig1}
\end{figure}

\section{Experimental Method}

A scanning electron microscope image of a device similar to the one used in this study is shown in Fig. 1a. The device is fabricated on a $GaAs/Al_{x}Ga_{1-x}As$ wafer using wet etching to define two conducting channels (running from top to bottom in the figure) and electron-beam lithography to define metallic gates above the channels. The right-hand channel of the device is an electron pump similar to ones used in previous studies \cite{giblin2012towards}. Two metallic gates termed the entrance and exit gates supplied with voltages $V_{\text{G1}}$ and $V_{\text{G2}}$ deplete the 2-DEG and form a quantum dot (QD) in the circular cut-out region between the gates. The left-hand channel contains a point-contact charge detector. When the channel conductance is set close to pinch-off by applying a negative voltage $V_{\text{PC}}$ to the point-contact gate, the conductance of the channel is a sensitive probe of the local charge environment \cite{field1993measurements}. To maximise the sensitivity of the PC to the QD charge state, the QD is offset from the centre of the pump channel towards the PC. 

In this work, the pump was operated in two modes. In the continuous pumping mode, the entrance gate was driven with an AC signal at frequency $f$ super-imposed on a DC voltage: $V_{\text{G1}}(t)=V_{\text{G1DC}}+Asin(2\pi f t)$, and the exit gate was supplied with a DC voltage $V_{\text{G2}}=V_{\text{G2DC}}$. This has the effect of pumping electrons from source (S) to drain (D), and generating a current $I_{\text{P}}$. In electron trapping mode, the PC channel was biased with a voltage $V_{\text{B}}=1$~mV, and $V_{\text{PC}}$ was tuned to maximise $dI_{\text{PC}}/dV_{\text{PC}}$. The pump gates were driven with the voltage sequence illustrated in Fig. 1b. At time $t=0$, the entrance gate is pulsed to a positive voltage $V_{\text{G1L}}$. This lowers the potential barrier under the gate and couples the QD to the source lead. At time $t=1$~ms (red vertical dashed line) the entrance gate is set to a negative voltage $V_{\text{G1M}}$ (the inset shows the rise time of this pulse). Depending on the setting of the exit gate $V_{\text{G2L}}$, the negative-going switch of the entrance gate at $t=1$~ms may load electrons into the QD and possibly also eject some electrons to the drain, leaving the QD containing $N$ electrons. The charge state $N$ is probed by measuring the PC current for a time $t_{\text{m}}$ to yield a value $I_{\text{PC1}}$. A wait time $t_{\text{w}}$ ensures that $I_{\text{PC1}}$ is not affected by the transient current induced by the entrance gate pulse. The pump gates couple strongly to the PC, and the adjustment of the entrance gate to $V_{\text{G2M}}$ at $t=2$~ms is a practical convenience which allows the readout of $N$ as a function of $V_{\text{G2L}}$ without continually re-tuning the PC gate voltage. To reduce the effect of $1/f$ noise in $I_{\text{PC}}$, the current reading $I_{\text{PC1}}$ was referenced to a second PC current $I_{\text{PC2}}$ measured with a known QD state $N=0$. This state was obtained by setting the exit gate to a large negative value $V_{\text{G2H}}$ during the negative-going transition of the entrance gate (vertical dash-dot line at $t=28$~ms). The complete cycle of pump gate voltages yields a PC difference signal $\Delta I_{\text{PC}} = I_{\text{PC2}} - I_{\text{PC1}}$. For all the measurements reported here, $V_{\text{G1L}}=0.25$~V, $V_{\text{G2H}}=-0.9$~V, $t_{\text{w}}=10$~ms and $t_{\text{m}}=17$~ms. The measurements were performed at a temperature of $1.5$~K.

\begin{figure}
\includegraphics[width=8.5cm]{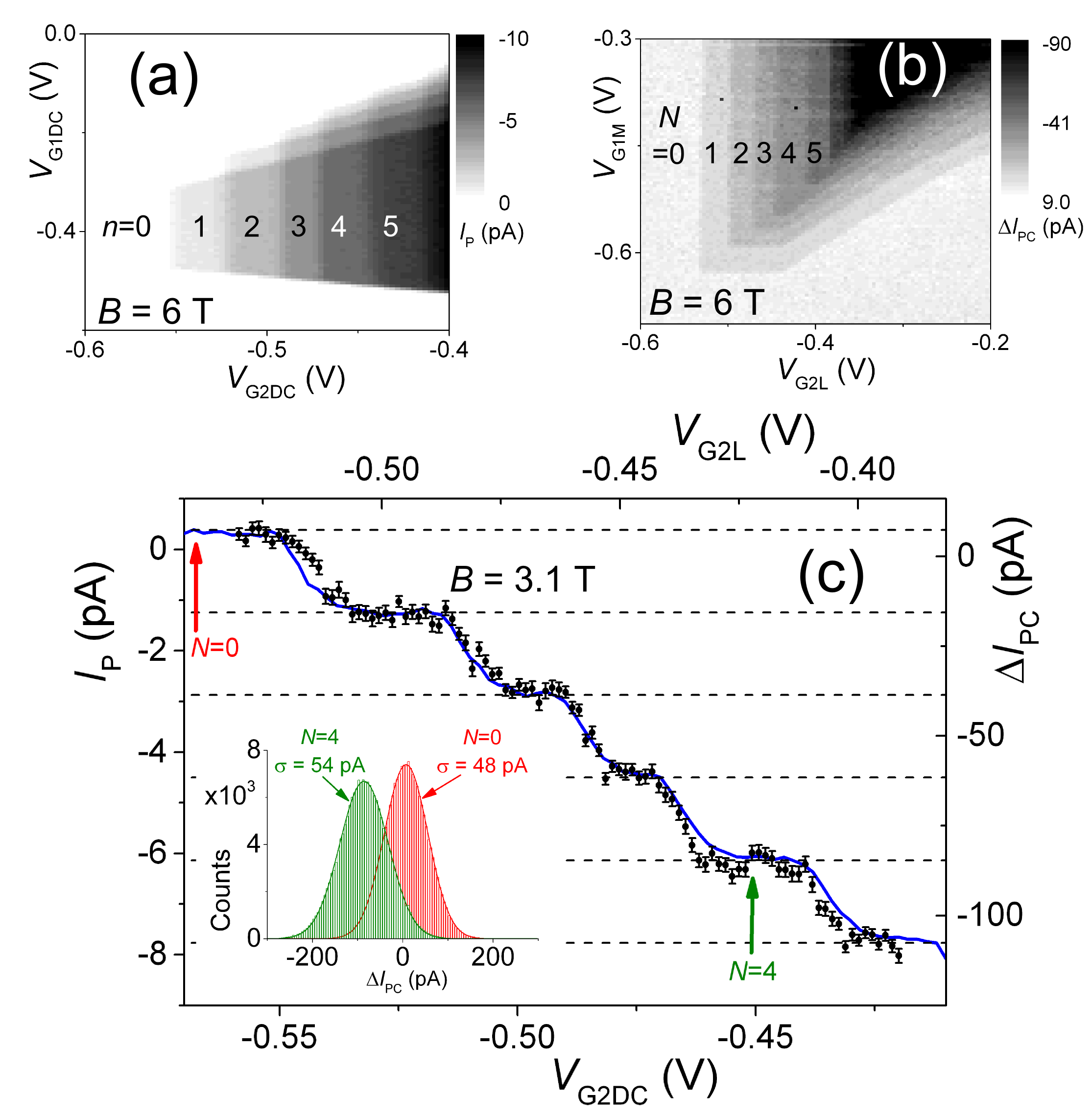}
\caption{\textsf{(a): Grey-scale map of the pumped current, with the device operating in continuous pump mode. $f=10$~MHz at RF amplitude $A=0.5$~V, $B=6$~T. The plateaus corresponding to $n$ pumped electrons per cycle for $0\leq n\leq5$ are indicated. (b): Grey-scale map of the PC difference signal with the device in trapping mode. $V_{\text{G2M}}=-0.4$~V. The plateaus corresponding to $N$ trapped electrons for $0\leq N\leq5$ are indicated. (c): Solid line, left and bottom axes: $I_{\text{P}}$ in pump mode, $f=10$~MHz $A=0.5$~V, $B=3.1$~T, $V_{\text{G1DC}}=-0.4$~V. Points, right and top axes: $I_{\text{PC}}$ in trap mode: $V_{\text{G2M}}=-0.4$~V, $V_{\text{G1M}}=-0.4$~V. Each point is averaged over 500 detection cycles, and the error bars show the standard error of the mean. Horizontal dashed lines indicate the PC signal corresponding to different numbers of trapped electrons. They are separated by $\Delta I_{\text{1e}}=23$~pA and offset by $+7$~pA. Inset: Histograms of $I_{\text{PC}}$ at the operating points indicated by the two arrows in the main figure. The solid lines are fits to Gaussian functions, with the standard deviations $\sigma$ of the fits indicated on the plots. The histograms are complied from $\sim 200000$ cycles at each operating point.}}
\label{fig:fig2}
\end{figure}

\section{Results and Discussion}

In Fig. 2(a) we show a grey-scale plot of $I_{\text{P}}$ as a function of $V_{\text{G1DC}}$ and $V_{\text{G2DC}}$, with the device in continuous pumping mode in a perpendicular magnetic field of $6$~T. The series of quantised current plateaus corresponding to $I_{\text{P}}=nef$, where n is the number of electrons pumped for each cycle of $f$, is a familiar characteristic of the tunable-barrier pump. Fig. 2(b) shows the corresponding map of $\Delta I_{\text{PC}}$ with the device in trapping mode, and  $V_{\text{G2M}}=-0.4$~V. Each pixel in the plot results from averaging $\Delta I_{\text{PC}}$ over 100 cycles of the type shown in Fig. 1b. The plot was constructed from scans of $V_{\text{G2L}}$ with constant $V_{\text{G1M}}$ and $V_{\text{PC}}$. Bands corresponding to a constant number $N$ of trapped electrons are clearly visible. As $V_{\text{G2L}}$ is increased from a large negative value, $N$ exhibits a step-wise increase, with the step position independent of $V_{\text{G1M}}$. This is to be expected, since $V_{\text{G1M}}$ determines the energy above the Fermi level that the electrons are held, but does not affect the loading dynamics. If $V_{\text{G2L}}$ is increased sufficiently, $N$ decreases in a series of steps back down to zero. This is because the exit barrier is no longer high enough to maintain all the loaded electrons in the dot, and electrons are ejected to the drain \cite{mcneil2011demand}. The truncation of the constant $N$ steps at large negative $V_{\text{G1M}}$ is an artefact caused by the choice of $V_{\text{G2M}}$. For $V_{\text{G2M}} > V_{\text{G2L}}$ electrons can be ejected to the drain during the step of $V_{\text{G2}}$ at $t=2$~ms. In the subsequent experiments reported here, we ensured that $V_{\text{G1M}}$ was sufficiently positive to avoid this artifact, and we verified that the results were independent of the choice of $V_{\text{G2M}}$.

In Fig. 2c we compare the step-like increase of both $n$ and $N$, with similar operating conditions at $B=3.1$~T. The entrance gate voltage in pump mode has been tuned so that all the electrons which are loaded from the source are also ejected to the drain. The x-axes of the plots have been shifted to bring them into alignment, but the scaling factor is the same. The overlap between the two types of data emphasises that the two modes of operation of the device are probing the same loading dynamics. Comparing the data of Fig. 2(b) and (c), it is apparent that the PC signal corresponding to one extra electron in the QD, $\Delta I_{\text{1e}}$ is field-dependent. An offset $\sim 7$~pA in $\Delta I_{\text{PC}}$ is visible in the data of Fig. 2(c). This is believed to be due to activation of trap states close to the PC (the offset in $I_{\text{P}}$ is a trivial pre-amp offset). The overall noise level in $\Delta I_{\text{PC}}$ is illustrated by the histograms of $\Delta I_{\text{PC}}$, obtained at two values of $V_{\text{G2L}}$ corresponding to $N=0$ and $N=4$ (inset). It is clear from the width of the histograms that single-shot measurement of the charge state $N$ in one loading cycle is not possible in this sample. The different standard deviations obtained from Gaussian fits to the histograms are due to slightly different $V_{\text{PC}}$. We will return to discuss the noise and detection fidelity in a later section of this paper.

\begin{figure}
\includegraphics[width=8.5cm]{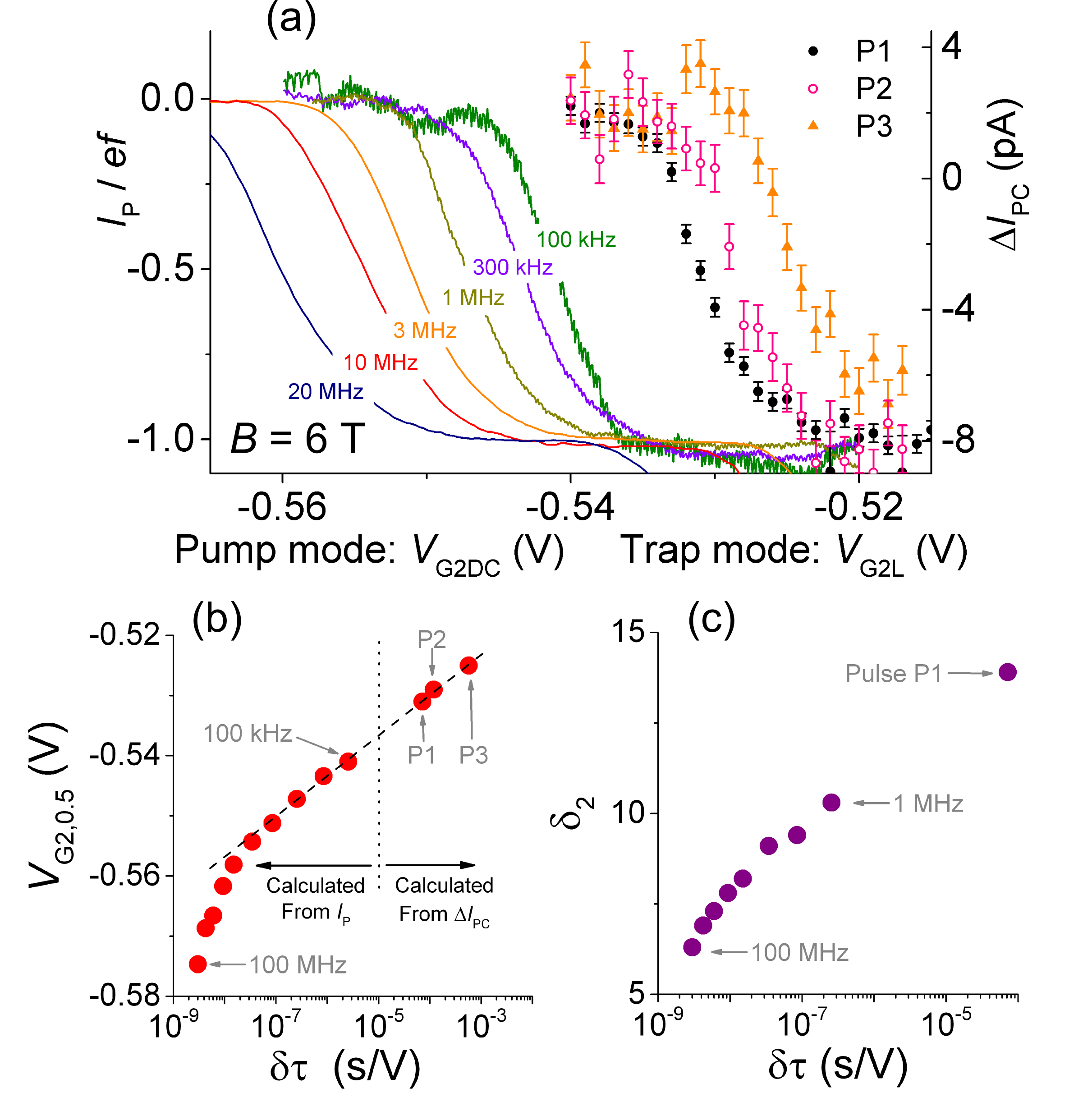}
\caption{\textsf{(a): Solid lines, left axis: $I_{\text{P}}$ normalised to $ef$ as a function of $V_{\text{G2DC}}$ for a range of pump frequencies at $B=6$~T. $A=0.5$~V for $f\geq 10$~MHz and $0.65$~V for $f< 10$~MHz. Points, right axis: $\Delta I_{\text{PC}}$ as a function of $V_{\text{G2L}}$ with $V_{\text{G1M}}=-0.4$~V. $V_{\text{G2M}}=-0.52$~V for data $P1$ and $-0.53$~V for data $P2,P3$. The labels $P1-P3$ refer to different pulse rise times. (b): Shift in plateau position along the $V_{\text{G2}}$ axis as a function of entrance barrier rise time. The error bars are smaller than the data points and the dashed line is a guide to the eye. (c): delta-2 parameter from fit to decay cascade model as a function of entrance barrier rise time.}}
\label{fig:fig3}
\end{figure}

In the data of Fig. 3 we focus on the transition from loading zero electrons, to loading one electron into the QD. Fig 3a (solid lines,left axis) shows the pumped current with the pump driven at a range of relatively low frequencies. As the frequency is reduced, the plateau shifts to more positive exit gate voltage, and the width of the transition is slightly reduced. The current also becomes difficult to measure with the conventional ammeter used to measure $I_{\text{P}}$ in this experiment ($ef=16$~fA for $f=100$~kHz). Data points (right axis) show $\Delta I_{\text{PC}}$ for three different entrance gate pulse rise-times, denoted $P1-P3$. $P1$ is as illustrated in the inset to Fig. 1(b), and $P2$ and $P3$ are progressively more heavily filtered with longer rise times. It has been predicted theoretically using the decay-cascade model \cite{fujiwara2008nanoampere,kashcheyevs2010universal}, and verified experimentally over a limited range \cite{fujiwara2008nanoampere} that the plateaus positions shift to more positive exit gate voltage in proportion to $log(\tau_{rise})$ where $\tau_{rise}$ is the time-scale for raising the entrance barrier. $\tau_{rise}$ is proportional to $\frac{1}{-(dV_{\text{G1}}/dt)}$ at the point in the $V_{\text{G1}}(t)$ cycle where the one-electron dot level rises above the Fermi level. We have defined $\delta\tau =\frac{1}{-(dV_{\text{G1}}/dt|_{\text{VG1=-0.2 V}})}$. The choice of $V_{\text{G1}}=-0.2$~V is somewhat arbitrary, but the following results are almost completely independent of this choice in the plausible range of $V_{\text{G1}}$ from $0$~V to $-0.3$~V. We note that Fig. 2(b) shows that the QD is isolated from the source lead and containing a stable number of electrons for $V_{\text{G1}}\geq -0.3$~V. In Fig. 3(b) we show the shift of the plateau as a function of $\tau_{rise}$, where the shift is quantified as the exit gate voltage $V_{\text{G2,0.5}}$ for which $n=0.5$ (pump data) or $N=0.5$ (trap data). Our measurement of $N$ using relatively slow loading pulses allows us to verify the expected shift in plateau position over four orders of magnitude in $\delta\tau$. 

For $f\gtrsim 10$~MHz, we see deviation of the data of Fig. 3(b) from straight-line behaviour. This is due to significant distortion of the $I_{\text{P}}(V_{\text{G2DC}})$ characteristic . The decay-cascade model only predicts a shift in the plateau along the $V_{\text{G2}}$ axis, but not any change in the shape of the plateau as a function of the pumping time-scale. We fitted some of the data sets in Fig. 3(a) to the decay-cascade formula $\frac{I_{\text{P}}}{ef}=exp(-exp(-\alpha V_{\text{G2}} + \Delta_{1})) + exp(-exp(-\alpha V_{\text{G2}} + \Delta_{2}))$ in the range $0 \leq \frac{I_{\text{P}}}{ef} \leq 1.5$ and extracted the parameter $\delta_{2} = \Delta_{2} - \Delta_{1}$, plotted in Fig. 3(c). A similar decrease in $\delta_{2}$ with increasing pumping frequency has been observed previously \cite{giblin2012towards}, but at much higher frequencies. The observation of this trend at relatively low frequencies, where capacitive cross-talk between the pump gates is expected to be negligible, is strong evidence that the decay-cascade model does not capture some key features of the pumping mechanism. The decoupling of a quantum dot from a reservoir by a rising tunnel barrier is a problem of general theoretical interest. More recent work includes extensions of the decay cascade model to include the effect of finite temperature in the leads \cite{fricke2013counting,yamahata2014accuracy}, and a non-Markovian model including the energy scales associated with the time-dependence of the barrier height \cite{kashcheyevs2012quantum}. Detailed comparisons of the predictions of these models with experimental data, as a function of barrier rise time, will be the subject of future research.

\section{Simulation}

We now consider whether single-shot detection of the QD charge state is possible in this type of experimental geometry. In Fig. 4(a) we show the noise spectrum of $I_{\text{PC}}$ for the operating conditions of Fig.2c. This spectrum has a strong $1/f$ characteristic, which is commonly associated with the fluctuation of an ensemble of charged defects. For comparison we show the noise spectrum of the current pre-amplifier connected to a dummy load designed to simulate the electrical impedance of the PC and its associated wiring ($0.5$~nF in parallel with $100$~k$\Omega$). It is likely that the PC in our experiment exhibits excess noise because it is formed at the edge of a chemical-etched channel which contains a high density of charged defects. Alternative device designs can form the PC using electrostatic gates only, and in the following discussion we will assume that the experimental noise will be dominated by the loaded pre-amp noise. We used a numerical simulation to generate noise with a similar frequency spectrum to the loaded pre-amp (black line in Fig. 4a) and then used this noise as an input to a simulation of the experimental protocol illustrated in Fig. 1b. For a given $t_{\text{W}}$ and $t_{\text{M}}$ the output of the simulation is a histogram of $\Delta I_{\text{PC}}$ (points in Fig. 4b), which we fit to a Gaussian function (solid line in Fig. 4b). The ability of the experiment to distinguish charge states differing by one electron depends on the width $\sigma$ of this histogram, and the sensitivity of $\Delta I_{\text{PC}}$ to a change in $N$ of one, denoted $\Delta I_{\text{1e}}$. In Fig. 4b we illustrated the $N=1$ state, for the case where $\Delta I_{\text{1e}}=46$~pA, i.e. twice the experimentally measured value. We define a threshold current $\Delta I_{\text{th}}$, and assign the state $N=0$ or $N=1$ to measurements with $\Delta I_{\text{PC}}<\Delta I_{\text{th}}$ and $\Delta I_{\text{PC}}>\Delta I_{\text{th}}$ respectively. The probability $P_{\text{err}}$ of assigning the wrong charge state to a measurement result is simply the integral of $P_{\text{N=1}}$ from $\Delta I_{\text{PC}}=\Delta I_{\text{th}}$ to infinity, where $P_{\text{N=1}}$ is the normalised probability distribution of obtaining a value $\Delta I_{\text{PC}}$ in a single measurement cycle. To ensure $P_{\text{err}}<10^{-8}$, as is generally needed for primary electrical metrology, we require $\Delta I_{\text{1e}}=11.2\sigma$. Thus, for $\sigma =9.7$~pA obtained in the simulation, we require $\Delta I_{\text{1e}}=109$~pA, or more than four times the measured value of $23$~pA. Planar QD-PC systems with optimised geometry have demonstrated $\Delta I_{\text{1e}}\sim 300$~pA with $V_{\text{B}}=1$~mV \cite{vandersypen2004real}, and sensitivities an order of magnitude larger have been demonstrated with vertical QD-PC geometry \cite{gustavsson2008measuring}.

\begin{figure}
\includegraphics[width=8.5cm]{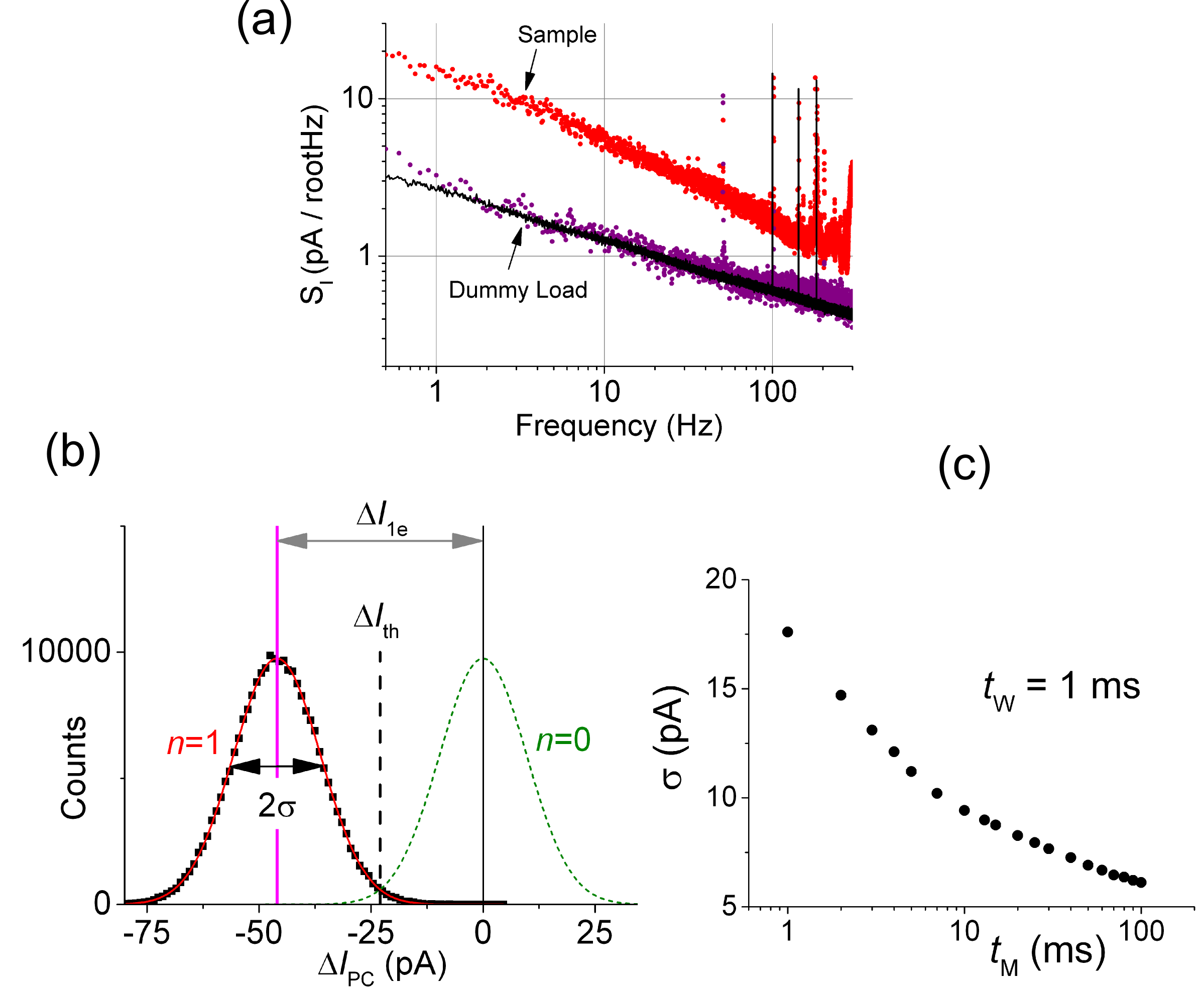}
\caption{\textsf{(a): Spectra of noise in $I_{\text{PC}}$ at $B=3.1$~T (filled circles, labeled 'sample') compared to the noise of the current preamp connected to a dummy load simulating the electrical impedance of the PC as seen by the preamp (filled squares, labeled 'dummy load'). The solid line overlapping the dummy load data is the spectrum of simulated time-domain data with the same characteristics as the pre-amp. (b): filled squares: histogram of $\Delta I_{\text{PC}}$ calculated in a simulation of the trapping mode of the experiment. Solid line: Gaussian fit to the points with standard deviation $\sigma$. Dotted line: the same Gaussian, shifted by $+46$~pA to have its centre at $\Delta I_{\text{PC}}=0$~pA. (c): Standard deviation $\sigma$ as a function of measurement time $t_{\text{m}}$ for simulated experimental runs with fixed $t_{\text{w}}=1$~ms.}}
\label{fig:fig3}
\end{figure}

Finally, we consider the possibility of completing the required large number of test cycles in a reasonable time-scale. For $t_{\text{W}}=8$~ms, $t_{\text{M}}=13$~ms, $10^8$ cycles would take $49$~days. By shortening the measurement and wait times, more test cycles can be completed at the expense of a lower detection fidelity. Fig 4c shows $\sigma$ for simulated data, with $t_{\text{W}}=1$~ms and variable $t_{\text{M}}$. For $t_{\text{M}}=1$~ms, one cycle takes $4$~ms and $2.16 \times 10^7$ cycles could be completed in $24$ hours, although $\sigma$ is now $17$~pA, requiring  $\Delta I_{\text{1e}}=190$~pA in order to maintain the detection fidelity at the $10^{-8}$ level. The overall cycle time could be shortened further, because for $t_{\text{W}}\lesssim 5$~ms, $\sigma$ is determined mainly by frequency-independent noise and it is not necessary to perform an $N=0$ reference cycle for every loading cycle. Using $t_{\text{W}}=1$~ms in the simulations assumes the use of a current pre-amplifier with a shorter settling time than the one used in the experiments reported here, but this is well within the achievable specifications for room-temperature  \cite{kretinin2012wide} and cryogenic \cite{vink2007cryogenic} current pre-amplifiers. 

\section{conclusions}

Using the methodology presented in this paper, metrological error-detection on the tunable barrier pump is feasible with modest improvements to the PC-QD coupling sensitivity, and the bandwidth of the current pre-amplifier used to measure the PC current. Using an arbitrary waveform generator to drive the pump entrance gate, the loading and ejection errors can be investigated separately. More generally, the initialisation of a dynamic quantum dot can be investigated over a very wide range of the decoupling rate of the dot from the leads, including slow rates not accessible by measuring the pumped current. This should improve our ability to distinguish between different models for the dot initialisation process.

\begin{acknowledgments}
This research was supported by the UK department for Business, Innovation and Skills and within the Joint Research Project "`Quantum Ampere"' (JRP SIB07) within the European Metrology Research Programme (EMRP). The EMRP is jointly funded by the EMRP participating countries within EURAMET and the European Union.
\end{acknowledgments}



\bibliography{SPGrefsEDet}
\bibliographystyle{apsrev}

\end{document}